\documentclass[12pt]{spieman}  
\usepackage{amsmath,amsfonts,amssymb}
\usepackage{graphicx}
\usepackage{setspace}
\usepackage{tocloft}
\usepackage{xspace}
\usepackage[euler]{textgreek}

\newcommand{\mum}{\mbox{{\usefont{U}{eur}{m}{n}{\char22}}m}\xspace}

\newcommand{\sce}{\mbox{SCE\lowercase{x}AO}\xspace}

\title{Characterizing Vibrations at the Subaru Telescope for the Subaru Coronagraphic Extreme Adaptive Optics instrument}

\author[a,*]{Julien Lozi}
\author[a,b,c,d]{Olivier Guyon}
\author[e]{Nemanja Jovanovic}
\author[a]{Naruhisa Takato}
\author[f]{Garima Singh}
\author[g]{Barnaby Norris}
\author[a]{Hirofumi Okita}
\author[a]{Takamasa Bando}
\author[h]{Frantz Martinache}
\affil[a]{National Astronomical Observatory of Japan, National Institutes of Natural Sciences, Subaru Telescope, 650 North A$\!$`oh\={o}k\={u} Place, Hilo, HI 96720, U.S.A.}
\affil[b]{Steward Observatory, University of Arizona, Tucson, AZ 85721, U.S.A.}
\affil[c]{College of Optical Sciences, University of Arizona, Tucson, AZ 85721, U.S.A.}
\affil[d]{Astrobiology Center of NINS, 2-21-1, Osawa, Mitaka, Tokyo, 181-8588, Japan}
\affil[e]{Department of Astronomy, California Institute of Technology, Pasadena, CA 91125, U.S.A.}
\affil[f]{Jet Propulsion Laboratory, 4800 Oak Grove Drive, MS 183-901, Pasadena, CA 91109, U.S.A.}
\affil[g]{Sydney Institute for Astronomy, Institute for Photonics and Optical Science, School of Physics, University of Sydney, NSW 2006, Australia}
\affil[h]{Observatoire de la C\^{o}te d'Azur, Boulevard de l'Observatoire, 06300 Nice, France}

\cftpagenumbersoff{figure}
\cftpagenumbersoff{table} 
\begin{document} 
\maketitle

\begin{abstract}

Vibrations are a key source of image degradation in ground-based instrumentation, especially for high-contrast imaging instruments. Vibrations reduce the quality of the correction provided by the adaptive optics system, blurring the science image and reducing the sensitivity of most science modules. We studied vibrations using the Subaru Coronagraphic Extreme Adaptive Optics (\sce) instrument at the Subaru Telescope as it is the most vibration-sensitive system installed on the telescope. We observed vibrations for all targets, usually at low frequency, below 10~Hz. Using accelerometers on the telescope, we confirmed that these vibrations were introduced by the telescope itself, and not the instrument. It was determined that they were related to the pitch of the encoders of the telescope drive system, both in altitude and azimuth, with frequencies evolving proportionally to the rotational speed of the telescope. Another strong vibration was found in the altitude axis of the telescope, around the time of transit of the target, when the altitude rotation speed is below 0.12~arcsec/s. These vibrations are amplified by the 10-Hz control loop of the telescope, especially in a region between 4 and 6~Hz. In this work, we demonstrate an accurate characterization of the frequencies of the telescope vibrations using only the coordinates ---right ascension and declination--- of the target, and provide a means by which we can predict them for any telescope pointing. This will be a powerful tool that can be used by more advanced wavefront control algorithms, especially predictive control, that uses informations about the disturbance to calculate the best correction.

\end{abstract}

\keywords{Extreme Adaptive Optics, Vibrations, Control, Accelerometers, Encoders}

{\noindent \footnotesize\textbf{*}Julien Lozi,  \linkable{lozi@naoj.org} }

\section{Introduction}
\label{sec:intro}  

The strongest pointing disturbances faced by astronomical telescopes are atmospheric turbulence and mechanical vibrations. While turbulence is a random process with a power spectrum density decreasing at higher frequency proportional to a power law with a factor -11/3\cite{Kolmogorov1941}, mechanical vibrations can usually be described as a deterministic process at specific frequencies. 

Vibrations become a critical issue when adaptive optics (AO) systems are used, especially for high-contrast imaging. In this case, the extreme AO (ExAO) system is expected to deliver high quality images, with Strehl ratios (SR) over 80\%. Vibrations reduce the Strehl ratio by blurring the image over the course of a long exposure (10--60~s). If a coronagraph is used to mask the starlight to reveal fainter close-in companions, vibrations also create leakage around the edges of the coronagraph, reducing the starlight suppression level. High-contrast instruments such as the Subaru Coronagraphic Extreme Adaptive Optics (\sce)\cite{Jovanovic2015} instrument counteract this by using a wavefront control loop composed of a wavefront sensor (WFS) to measure the disturbance and a deformable mirror (DM) to control it. Despite the fact that the control loop of \sce runs at up to 3.6~kHz with sub-millisecond level latency, it has demonstrated a difficulty to correct low-frequency vibrations on occasion, due to the high amplitude and variability of the vibrations. Indeed, these high-amplitude vibrations can reach the limit of linearity of the WFS, or the limit of motion of the DM. Other high-contrast instruments are also impacted by vibrations, such as the Gemini Planet Imager (GPI)\cite{Hartung2014}, and the Spectro-Polarimetric High-contrast Exoplanet REsearch (SPHERE) instrument\cite{Sauvage2010}.

Since 1999, accelerometers were occasionally installed on the Subaru Telescope, and the telescope vibrations have been studied~\cite{Kanzawa2006}. This work presented some insights into the origin of some of the vibrations, and offered possible steps to reduce them. Unfortunately, no major improvements where made on the pointing of the telescope, and this is now one of the limiting factors in the performance of \sce. To better understand the origins of these vibrations, the same accelerometers where re-installed in 2015, to compare the telescope vibrations with the measurements taken from various wavefront sensors and imagers on board \sce . Here, we focus on the analysis of the vibrations measured with \sce, and compare them to the accelerometer data.  Methods to correct these vibrations will be presented in a separate publication. While a preliminary analysis of the data was presented~\cite{Lozi2016a}, we have made further enhancements to the data analysis and model fitting and the new and improved results are presented here. Section~\ref{sec:impact} describes how vibrations impact the data quality of the \sce instrument, while Sec.~\ref{sec:measure} outlines how the wavefront sensors of \sce and accelerometers installed on the telescope were used to measure these vibrations. Section~\ref{sec:telescope} presents an analytical analysis of the speed of the telescope during the tracking of a target, and finally Sec.~\ref{sec:analysis} compares this analysis to the data, showing that most of the vibrations degrading \sce's performances are originating from the telescope.

\section{IMPACT OF VIBRATIONS}
\label{sec:impact}

\subsection{Strehl Ratio as a Function of Exposure Time}
\label{sec:strehl}

The SR of the point-spread function (PSF), is a commonly used metric for the image quality of a high-contrast imager/adaptive optics system. The measurement of the SR is dependent on the exposure time of the acquisition. For example, a camera acquiring frames at a slower rate than a vibration will produce a blurred image with a reduced SR. On \sce, multiple modules can record focal plane images, from 600~nm to 2.3~\mum, at frequencies between 10~mHz to a few kHz\cite{Jovanovic2015}. Inside \sce, the PSF quality can be monitored with the internal Near-IR (NIR) camera (Raptor Photonics, OWL SW1.7 CL HS). This camera is equipped with a $320\times256$-pixel InGaAs detector, capable of acquiring full-frame images at up to 170~Hz. A H-band filter is inserted in front of the camera to analyze the image quality between 1.5 and 1.7~\mum. 

\begin{figure}
\begin{center}
\begin{tabular}{c}
\includegraphics[width=0.7\columnwidth]{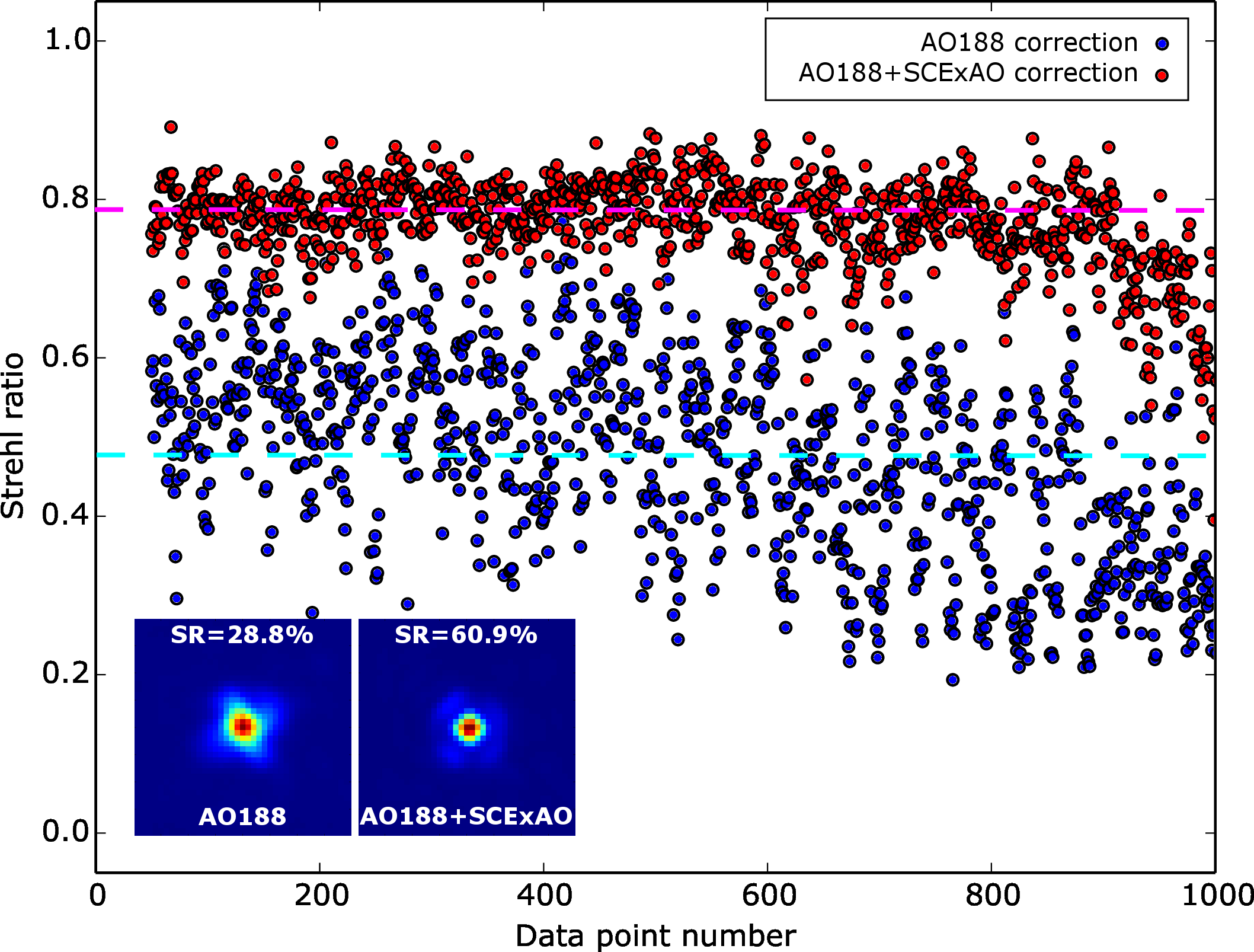}
\end{tabular}
\end{center}
\caption 
{ \label{fig:srvstime}
Evolution of the instantaneous Strehl ratio when the correction is provided by AO188, and AO188 and \sce combined. The data was taken in H-band at 170~Hz, over 1000~iterations ($\sim 6$~s). The average PSF of each case, simulated by averaging the 1000 5-ms exposures, is shown in the insert.} 
\end{figure}

The \sce instrument corrects the wavefront aberrations in two steps. \sce is placed at the output from Subaru Telescope's facility adaptive optics system, AO188. It benefits from the first stage correction of AO188, which delivers Strehl ratios between 20 and 40\% in the H-band in median seeing\cite{Minowa2010}. Using a Pyramid Wavefront Sensor (PyWFS) and a 2000-actuator deformable mirror, \sce performs the second stage of ExAO correction, bringing the SR up to 70--90\% in good seeing conditions\cite{Jovanovic2015}. Figure~\ref{fig:srvstime} presents the instantaneous on-sky Strehl ratio measured over 1000~frames ---or about 6~s at 170~Hz--- collected by the internal NIR camera with an exposure time of 5~ms, when only AO188 is correcting aberrations, and when AO188 and \sce are both operating. In both cases, the inserts in the lower left corner correspond to simulated long exposure images, obtained by averaging all the single 5-ms exposures. The two dashed lines correspond to the measured average Strehl ratio over the single exposures. In both case, a decrease in the Strehl ratio can be seen at the end of the time sequence, probably due to an increase of the turbulence level.

This figure shows that instantaneous SR measurements can vary greatly from frame-to-frame, especially when the correction is performed by AO188 alone. These variations are due to the residual atmospheric turbulence left uncorrected. In this case, the average instantaneous SR measurement (cyan dashed line) is 47.3\%, while the SR measured on the simulated long exposure ---the average of the 1000 5-ms exposures (left insert)--- is 28.8\%. When the ExAO correction from \sce is turned on, the instantaneous SR measurements increase significantly to an average of 78.6\% (magenta dashed line), and the deviation from this value is reduced, showing a better stability between images delivered by the ExAO system. However, the SR measured on the simulated long exposure (right insert) is only 60.9\%. This discrepancy between the average instantaneous SR and the SR of the simulated long exposure images is due to tip/tilt variations during the sequence. A fraction of the variations correspond to uncorrected atmospheric tip/tilt, but the majority are due to telescope and/or instrumental vibrations. In the example presented in Fig.~\ref{fig:srvstime}, the vibrations reduce the SR by about 20\%. Recent improvements in the wavefront control algorithm of \sce helped reduce the impact of vibrations, enabling us to reach SR during long exposures between 80 and 90\%.

\subsection{Effect of Vibrations on Interferometric Data in Visible}
\label{sec:interfero}

\sce is composed of several modules that analyze the light with different methods on various spectral channels, between 600~nm and 2.5~\mum. A few modules use interferometric methods to perform high contrast measurements, such as the Visible Aperture-Masking Polarimetric Interferometer for Resolving Exoplanetary Signatures (VAMPIRES), which combines non-redundant masking (NRM) interferometry and differential polarimetry to image the inner regions of debris disks and dust shells around stars\cite{Norris2015}.

\begin{figure}
\begin{center}
\begin{tabular}{c}
\includegraphics[width=0.5\columnwidth]{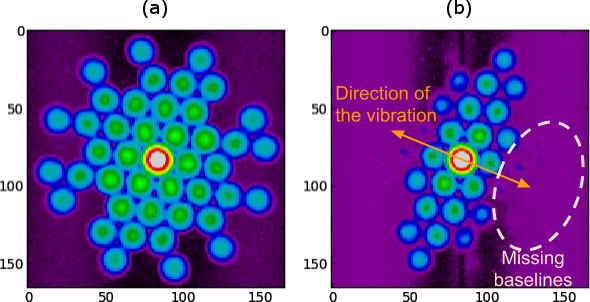}
\end{tabular}
\end{center}
\caption 
{ \label{fig:vampires}
MTF data from the polarimetric non-redundant masking mode of VAMPIRES in two different conditions: (a) when vibrations are low and (b) when vibrations are strong.} 
\end{figure}

Like classic imagers, interferometric imagers are equally affected by vibrations. Example images in different conditions were taken with VAMPIRES. Figure~\ref{fig:vampires} presents the on-sky modulation transfer function (MTF) derived from Fourier transform of focal plane images acquired with a 7-hole non-redundant mask (NRM) in the pupil plane, under two different sets of conditions: when the vibrations are low ($<100$~mas rms), and when they are very high ($>500$~mas rms). 

When vibrations are low in Fig.~\ref{fig:vampires} (a), the Fourier plane is fairly symmetrical circularly. All the baselines, defined by the spots in the figure, have visibilities significantly above the noise. When vibrations are strong ---in this case along a predominant direction shown by the arrow of Fig.~\ref{fig:vampires} (b)---, the extreme baselines are washed out and disappear in the noise in the direction of the vibration. This loss of signal at the longest baselines is consistent with a degradation of the resolution due to an elongation of the PSF.

Since the impact of vibrations on the image quality is inversely proportional to the wavelength, this effect is exacerbated in VAMPIRES because it operates at shorter wavelengths (600-800 nm)\cite{Jovanovic2015}. Thus, VAMPIRES and the other modules analyzing the visible light are therefore more sensitive to vibrations than their NIR counterparts.

These examples show that all the modules of the \sce are negatively affected by vibrations. In order to correct them efficiently it is essential to perform an accurate measurement and understand their origins.

\section{MEASURING INSTRUMENTAL AND TELESCOPE VIBRATIONS}
\label{sec:measure}

\subsection{Wavefront Sensing Metrology}
\label{sec:metrology}

The high speed telemetry of \sce's wavefront sensors can be used to analyze the vibrations degrading the PSF quality. During the time when the PyWFS was still in development, the coronagraphic Lyot-stop Low-Order Wavefront Sensor (LLOWFS) was used to track vibrations decomposed into tip and tilt\cite{Singh2014,Singh2015}. The LLOWFS uses the light rejected by the Lyot mask in the pupil plane of the coronagraph to measure relative changes in low-order Zernike modes, at speeds of up to 170~Hz, using the same model of camera as the internal NIR camera (see.Sec.~\ref{sec:strehl}).

\begin{figure}
\begin{center}
\begin{tabular}{c}
\includegraphics[width=0.5\columnwidth]{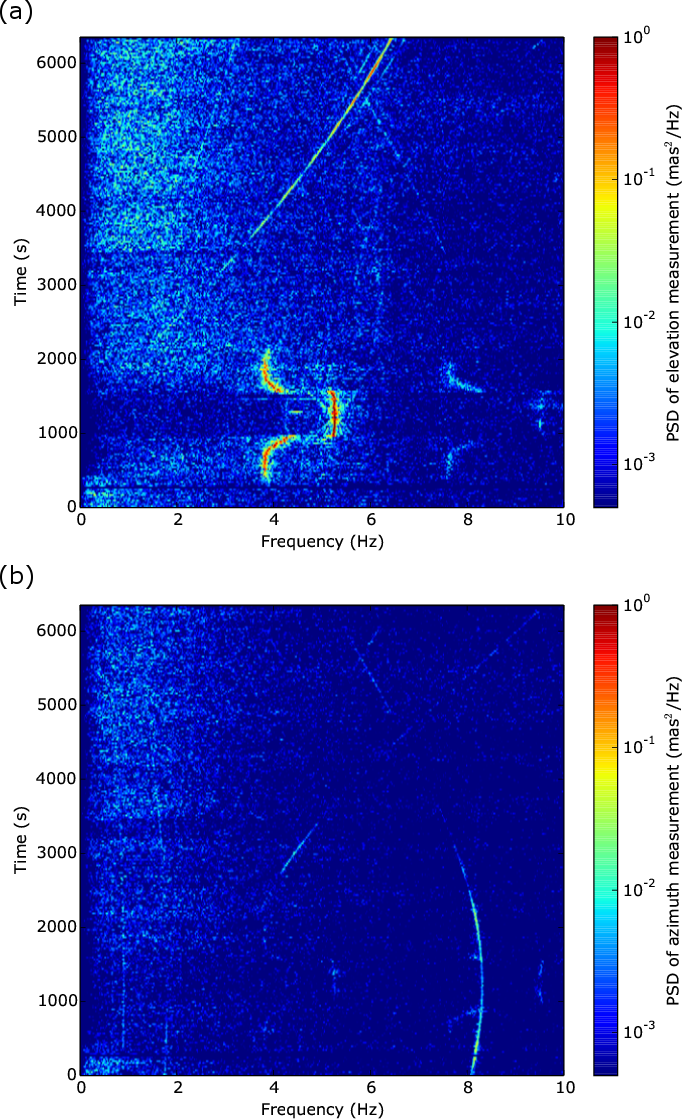}
\end{tabular}
\end{center}
\caption 
{ \label{fig:psdlowfs}
Evolution of the PSD of the pointing measurements provided by the Low-Order Wavefront Sensor, (a) in altitude and (b) in azimuth, for a target with a declination $\delta=47^o04'54"$. Transit of the target occurred at $t=1200$~s.}
\end{figure}

Since vibrations can evolve in frequency and amplitude with time, measurements are analyzed by looking at the evolution of the Power Spectrum Density (PSD) of a moving sample of usually 1000~iterations of the tip-tilt data. Figure~\ref{fig:psdlowfs} present such a result. These data were acquired by running the LLOWFS loop by itself without PyWFS correction, but after the correction from AO188, providing a Strehl ratio of about 30\%. In this case, the flux from the target was too low to run at full speed, the sampling frequency was therefore 20~Hz, and the moving window had a width of 50~s.

In high-contrast imaging, the image rotator ---a combination of mirrors  that rotates the image in any direction--- is used in fixed pupil mode, i.e. the pupil is fixed in the instrument while the field rotates. This mode keeps the speckles due to the diffraction of the telescope at the same position on the detector, while the target ---planet, disk, etc.--- rotates with the sky. Stellar speckles can then be subtracted more efficiently in post-processing using techniques like angular differential imaging (ADI)\cite{Marois2005, Lafreniere2007}. In the case of a fixed pupil mode, the direction of the altitude (or elevation) and the azimuth of the telescope are fixed with respect to the wavefront sensors. 

The tip-tilt data from the LLOWFS presented in Fig.~\ref{fig:psdlowfs} were rotated to match the altitude and azimuth axes of the telescope. In this figure, several vibrations are noticeable: in azimuth, some vibrations seem to evolve slowly in frequency and amplitude, while the altitude data show stronger vibrations, especially one evolving rapidly in frequency. The target crossed the local meridian at  $t=1200$~s, which explains the symmetry in the PSD for both axes about this time period. The frequencies and amplitudes of dominant vibration modes are related to the azimuth and elevation rotation speed of the telescope, therefore these vibrations can be attributed to the telescope more than to the instrument itself.

\begin{figure}
\begin{center}
\begin{tabular}{c}
\includegraphics[width=0.5\columnwidth]{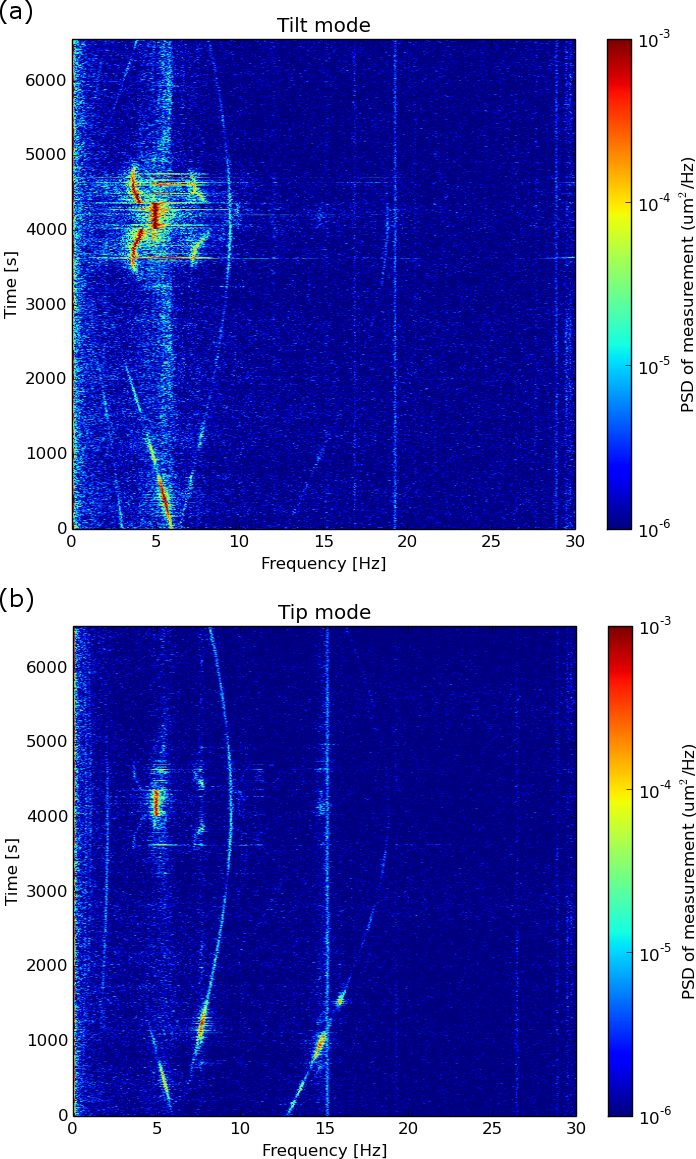}
\end{tabular}
\end{center}
\caption 
{ \label{fig:psdpywfs}
Evolution of the PSD of the pointing measurements provided by the Pyramid Wavefront Sensor, (a) in altitude and (b) in azimuth, for a target with a declination $\delta=43^o46'25"$. Transit of the target occurred at $t=4200$~s.} 
\end{figure}

The PyWFS was also used to study the vibrations. It is possible to track all the telescope and instrumental vibrations owing to the fact that the loop runs at 2~kHz routinely, and can be pushed up to 3.6~kHz. Since the loop corrects part of the vibrations in closed-loop operation, pseudo open-loop data are reconstructed from closed-loop measurements and the commands sent to the deformable mirror of \sce. This pseudo open-loop reconstruction excludes the first stage of correction performed by AO188, so these data simulate pointing errors left uncorrected by AO188, at the entrance of \sce. Figure~\ref{fig:psdpywfs} presents the frequency analysis of such a result, recorded during the science night of April 18, 2017. 

In this case, the data acquired at 2~kHz were averaged to simulate an acquisition frequency of 60~Hz, since most telescope vibrations are below 30~Hz. Similarly to the LLOWFS data, the PyWFS tip and tilt axes were rotated to match the altitude and azimuth axes of the telescope. Transit of the target occurred at $t=4200$~s. The patterns of vibrations measured with the PyWFS and the LLOWFS are very similar, with slow varying vibrations and harmonics on both axes, and a strong fast-varying vibration between 3.5 and 5~Hz, only around the transit, on the altitude axis. In this example, there are almost no vibrations after 15~Hz. This confirms that the main contributor of vibrations is the telescope itself.

\subsection{Accelerometers on the Telescope}
\label{sec:accelero}

\begin{figure}
\begin{center}
\begin{tabular}{c}
\includegraphics[width=0.5\columnwidth]{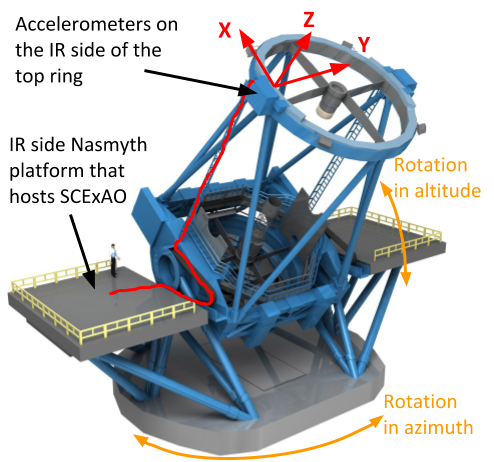}
\end{tabular}
\end{center}
\caption 
{ \label{fig:accelpos}
Position of the accelerometers on the IR side of the top ring of the Subaru Telescope. A mirror copy was also installed on the optical side of the top ring.} 
\end{figure}

To measure the motion of the telescope during on-sky operations, accelerometers were installed at the top of the telescope, on both sides of the ring that supports the secondary mirror. Figure~\ref{fig:accelpos} shows the position of the accelerometers installed on the Infrared (IR) side of the top ring. The red line represents the network cable running from the accelerometer acquisition unit and the SCExAO computer reading the measurements. 

The three accelerometers on the IR side are mounted at $90^o$ from each other. The X axis is tangential to the rotation in altitude of the telescope, the Y axis is radial to the top ring and the Z axis is parallel to the optical axis of the telescope. Contrary to the Gemini Planet Imager, \sce does not measure any strong vibration in focus\cite{Poyneer2016}, that would be generated by a variation in the distance between the primary and secondary mirrors. This is probably due to the fact that the structure holding the secondary mirror of Subaru (struts, top ring and spiders) is much stiffer than at Gemini Observatory, to accommodate heavy prime focus instruments. This was confirmed by accelerometer measurements in the Z-axis, that didn't feature any strong vibration. Therefore we will only present results with the X- and Y-axes in the rest of the paper. Similarly, since the second set of accelerometers is situated in the symmetric position on the other side of the top ring, their results are almost identical to the first set, and will be omitted in the rest of the paper for simplicity.

\begin{figure}
\begin{center}
\begin{tabular}{c}
\includegraphics[width=0.5\columnwidth]{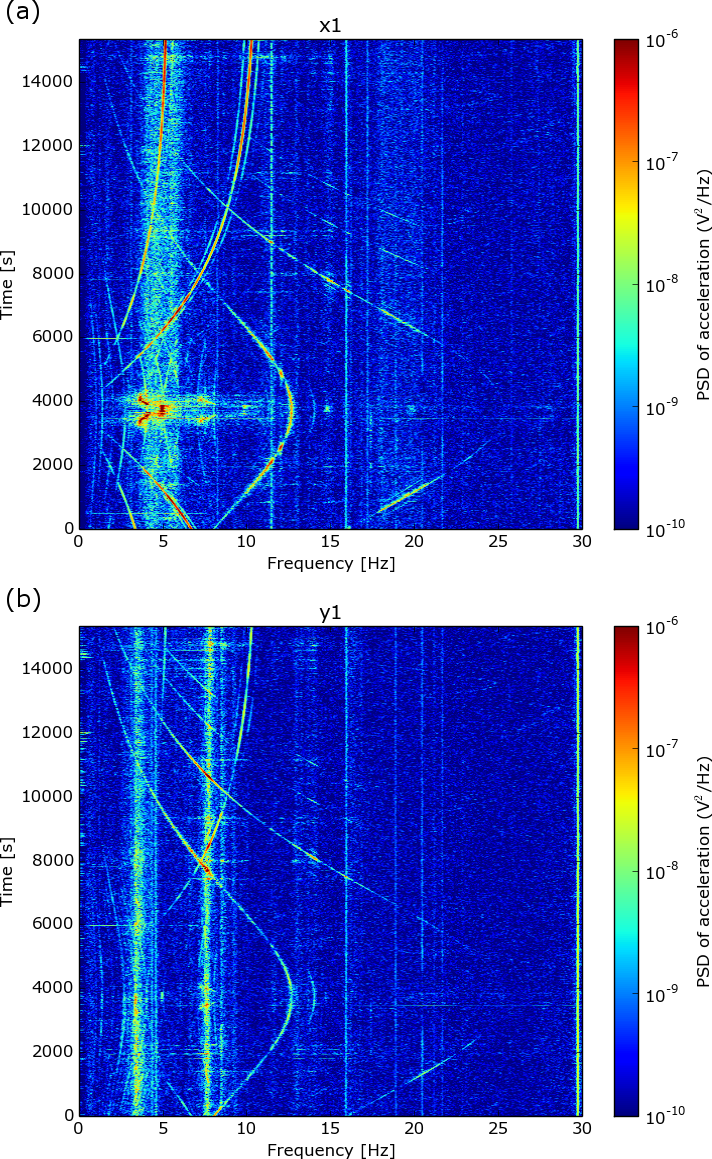}
\end{tabular}
\end{center}
\caption 
{ \label{fig:psdaccel}
Evolution of the PSD of accelerometer measurements, (a) in X and (b) in Y, for the target Vega (declination $\delta=38^o48'10"$). Transit of the target occurred at $t=3900$~s.} 
\end{figure} 

Figure~\ref{fig:psdaccel} presents an analysis of the accelerometer measurements in X and Y during the observation of the target Vega on the engineering night of August 12, 2016. The accelerometers were recorded at 60~Hz, although by design, their sensitivity drops around 20~Hz. In this example, the transit happened at $t=3900$~s. Once again, we see the similar pattern of vibrations as were presented in Fig.~\ref{fig:psdlowfs} and~\ref{fig:psdpywfs}, but at a position far from the \sce instrument, proving that the low-frequency vibrations seen by the instrument are introduced by the pointing of the telescope. We believe that the main eigenmode of the excited vibrations is similar to a solid body motion of the elevation axis, although it is possible that the top ring supporting the secondary mirror also moves relatively to the axis of the primary mirror, creating misalignment between the two. Another set of accelerometers will be installed close to the primary mirror to check this hypothesis. In Fig.~\ref{fig:psdaccel}, since the accelerometers cannot measure a pure altitude or azimuth motion, crosstalk can be seen as the same vibration streaks appearing in both axes. On the X axis in Fig.~\ref{fig:psdaccel} (a), we notice also a band of noise between 4 and 6~Hz, where some disturbances seem to be amplified. This is also where we can see the strong and fast-varying vibrations around the time of transit. On the Y axis in Fig.~\ref{fig:psdaccel} (b), a narrower band of vibration amplification is found at about 7.5~Hz, and at half that frequency, around 3.75~Hz.

The next two sections will attempt to explain the slow-varying vibrations observed in the sensors and the accelerometers, while Sec.~\ref{sec:transit} will analyze in more details the fast-varying vibration around transit.

\section{TEMPORAL ANALYSIS OF THE TELESCOPE MOTION}
\label{sec:telescope}

\subsection{Converting Equatorial Coordinates to Horizontal Coordinates}
\label{sec:conversion}

The position of a target in the sky is given by its right ascension $\alpha$ and declination $\delta$. For a telescope situated at a longitude $\lambda_0$ and latitude $\phi_0$, on any given night, the position of the target will be the same at the same local sidereal time $\theta_L$, given by
\begin{equation}
	\theta_L = \theta_G - \lambda_0 \, ,
\end{equation}
where $\theta_G$ is the Greenwich sidereal time. Once the local sidereal time is known, one can then convert the ascension coordinate into an hour angle $h$, such as
\begin{equation}
	h = \theta_L - \alpha = \theta_G - \lambda_0 - \alpha \, .
\end{equation}

If the telescope uses an altazimuth mount like the Subaru Telescope, the pointing angles in altitude ($a$) and azimuth ($A$) can then be expressed as
\begin{equation}
\label{eq:altaz}
\left\{
	\begin{array}{r@{\hspace{3pt}}l}
	\sin a &= \sin \phi_0 \sin \delta + \cos \phi_0 \cos \delta \cos h\\
    \tan A &= \dfrac{\sin h}{\cos h \sin \phi_0 - \tan \delta \cos \phi_0}
	\end{array}
\right.\, .
\end{equation}

Since the right ascension and the hour angle are usually described as a time, the conversion to radian necessary to compute Eq.~\ref{eq:altaz} is given by
\begin{equation}
	h[\mathrm{rad}] = \frac{2\pi}{T_\mathrm{sid}}h[\mathrm{sec}] \, ,
\end{equation}
where $T_\mathrm{sid} = 86164.1$~s is the duration of the sidereal day, or one full rotation of the night sky.

\begin{figure}
\begin{center}
\begin{tabular}{c}
\includegraphics[width=0.5\columnwidth]{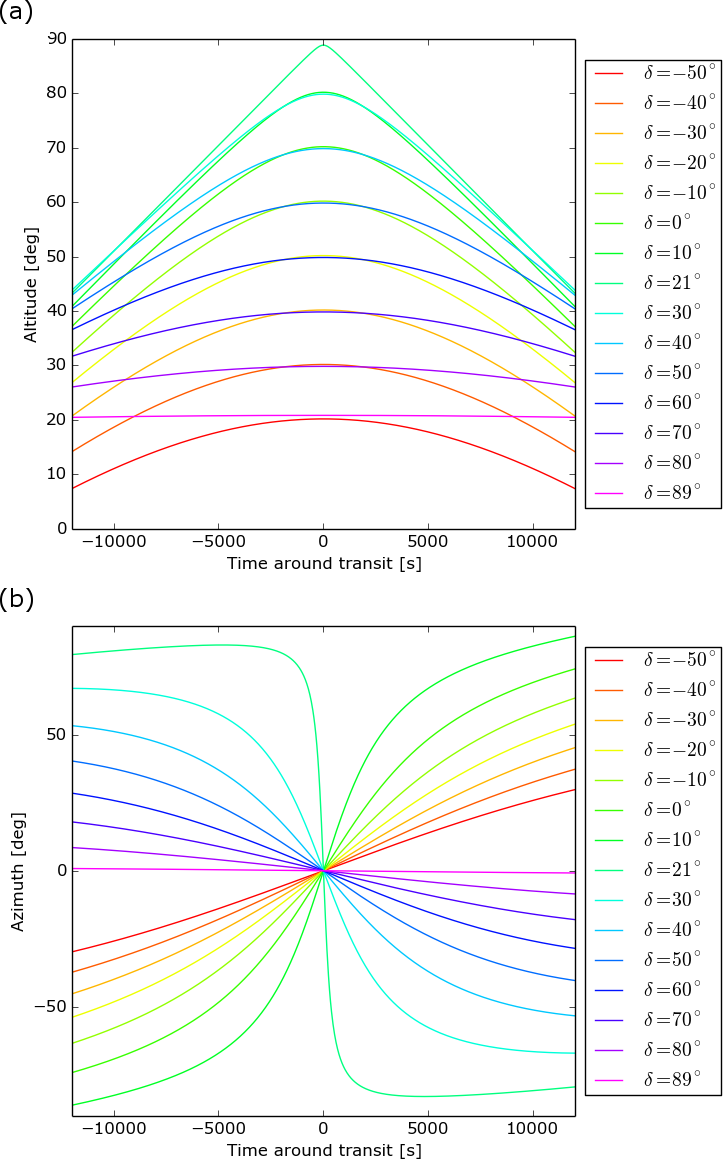}
\end{tabular}
\end{center}
\caption 
{ \label{fig:altazvsdel}
Position of the telescope in (a) altitude and (b) azimuth, while tracking a target with various $\delta$. } 
\end{figure} 

Figure~\ref{fig:altazvsdel} presents simulated pointing positions of the Subaru Telescope (latitude $\phi_0=19.8255^o$) in altitude and azimuth, for various declinations $\delta$, as a function of time around the transit of the target, when the hour angle $h=0$. For targets with declinations close to the latitude of the telescope $\phi_0$, the transit happens close to zenith, creating this almost triangular evolution of the altitude, while the azimuthal rotation goes through a rapid change. At the other extreme case, for targets close to the North pole, the altitude and azimuth of the telescope are almost constant.

\subsection{Rotational Speed of the Telescope During Tracking}
\label{sec:speed}

Since the hour angle $h$ is the only parameter varying with time in Eq.~\ref{eq:altaz}, then the rotational speed of the telescope can be calculated with the time derivative of that equation. The derivative of the altitude is then
\begin{equation}
\frac{\mathrm d a}{\mathrm d h} = \frac{-\cos \phi_0 \cos \delta \sin h}{\sqrt{1-\left(\sin \phi_0 \sin \delta + \cos \phi_0 \cos \delta \cos h\right)^2}}\, ,
\label{eq:derivea}
\end{equation}
and the derivative of the azimuth is
\begin{equation}
\frac{\mathrm d A}{\mathrm d h} = \frac{\sin \phi_0 - \tan \delta \cos \phi_0 \cos h}{\sin^2 h + \left(\cos h \sin \phi_0 - \tan \delta \cos \phi_0\right)^2}\, .
\end{equation}

The rotational speeds of the telescope in altitude ( $\omega_a$) and azimuth ($\omega_A$) are then given by
\begin{equation}
\omega_a = \left|\frac{\mathrm d a}{\mathrm d h}\right|\frac{2\pi}{T_\mathrm{sid}} \;\textrm{and}\; \omega_A = \left|\frac{\mathrm d A}{\mathrm d h}\right|\frac{2\pi}{T_\mathrm{sid}}\, .
\end{equation}

\begin{figure}
\begin{center}
\begin{tabular}{c}
\includegraphics[width=0.9\columnwidth]{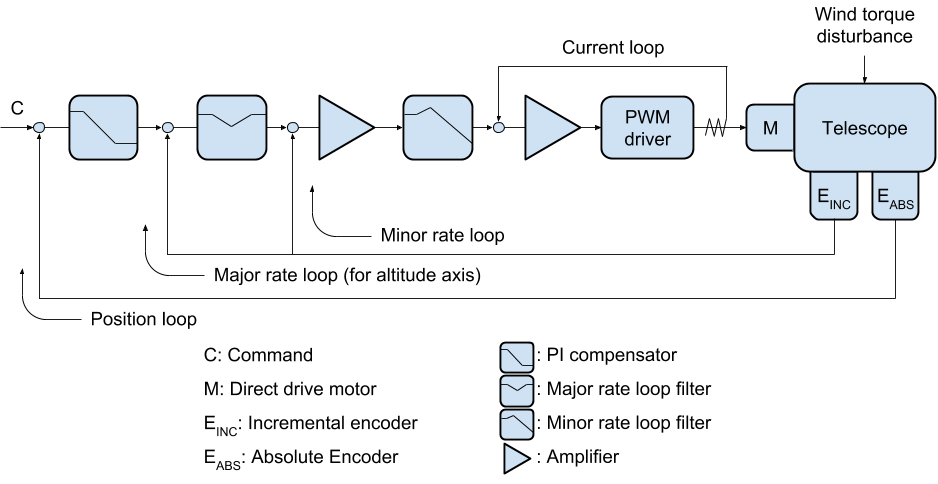}
\end{tabular}
\end{center}
\caption 
{ \label{fig:loop}
Block diagram of the control loop for azimuth and altitude axes (figure from \cite{Kanzawa2006}).} 
\end{figure}

Each axis of the telescope is equipped with two sets of encoders ---tapes with a regular pattern printed on them, read optically--- measuring the position of the telescope at any moment. One set performs an absolute but coarse measurement of the position, while the second performs a more precise incremental measurement of the rotational velocity of that axis\cite{Kanzawa2006}.  Figure~\ref{fig:loop} presents a block diagram of the servo control loop for both axes of the Subaru Telescope\cite{Nogushi1998}. The position loop uses the absolute encoders to measure the current position for each axis. Then the rate loops use the incremental encoders to measure speed.  A Pulse-Width Modulator (PWM) driver is used to regulate the voltage sent to the motors of the telescope, to increase the accuracy between two steps of the encoders at a sub-arcesecond per second level. The two absolute encoders have the same pitch of $p_\mathrm{abs} = 26.45$~arcsec, while the incremental encoders in altitude ($p_{a,\mathrm{inc}}$) and azimuth ($p_{A,\mathrm{inc}}$) have different pitches, with $p_{a,\mathrm{inc}} = 2.23$~arcsec and $p_{A,\mathrm{inc}} = 2.95$~arcsec.

For each axis and each encoder, the frequency corresponding to the pitch of the encoder as a function of rotation speed is given by
\begin{equation}
	f_{\mathrm{axis},\mathrm{encoder}} = \frac{\omega_\mathrm{axis}}{p_{\mathrm{axis},\mathrm{encoder}}}\, .
\label{eq:freq}
\end{equation}

This equation is essential for vibration analysis, because although the encoders are not creating vibrations on the telescope, a misalignment of the encoder heads or other misbehavior in the servo loops can create periodic errors in the pointing similar to mechanical vibrations, at the frequency of the pitch of the encoder, or potentially some harmonics of that frequency.

\begin{figure}
\begin{center}
\begin{tabular}{c}
\includegraphics[width=0.5\columnwidth]{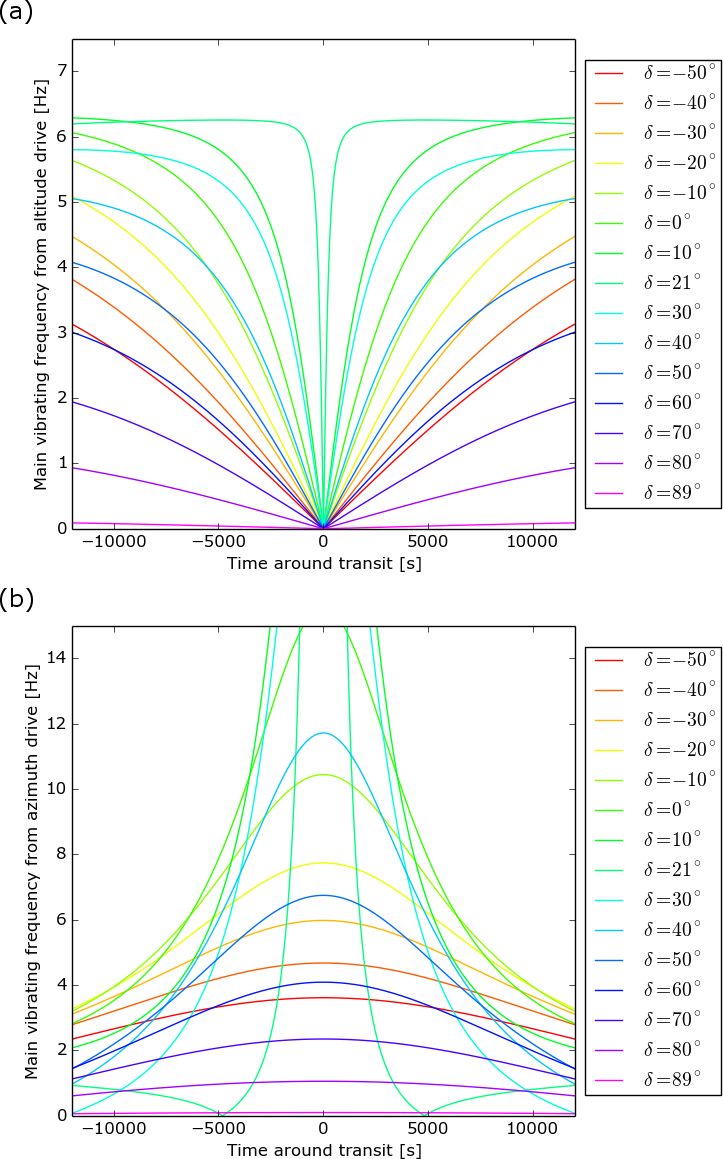}
\end{tabular}
\end{center}
\caption 
{ \label{fig:vibaltazvsdel}
Frequency of encoder reading (a) altitude and (b) azimuth, while tracking a target with various $\delta$. } 
\end{figure}

Using Eqs.~\ref{eq:derivea} to~\ref{eq:freq}, Fig.~\ref{fig:vibaltazvsdel} presents the evolution of $f_{a,\mathrm{inc}}$ and $f_{A,\mathrm{inc}}$ around transit, for different declinations $\delta$. 

In most cases, the frequencies calculated for the different targets are between 0 and 7~Hz, and their profiles are consistent with the vibrations measured in Fig.~\ref{fig:psdlowfs} and Fig.~\ref{fig:psdpywfs}. During the transit of the target, the frequency goes through zero in altitude, while it goes through its maximum in azimuth, defined by
\begin{equation}
	f_{A,\mathrm{inc},\mathrm{max}} = \frac{\left|\sin \phi_0\right|}{\left(\sin \phi_0 - \tan \delta \cos \phi_0\right)^2}\frac{2\pi}{T_\mathrm{sid}p_{A,\mathrm{inc}}} \, .
\end{equation}

From this equation and Fig.~\ref{fig:vibaltazvsdel}, we can see that for targets with declinations close to the latitude of the telescope, i.e. targets transiting at zenith, the speed of the telescope ---and therefore the frequency of vibrations--- undergoes more abrupt changes around zenith for both axes.

\section{FREQUENCY ANALYSIS}
\label{sec:analysis}

\subsection{Correlation Between Accelerometer and WFS Measurements}
\label{sec:correlation}

\begin{figure}
\begin{center}
\begin{tabular}{c}
\includegraphics[width=0.5\columnwidth]{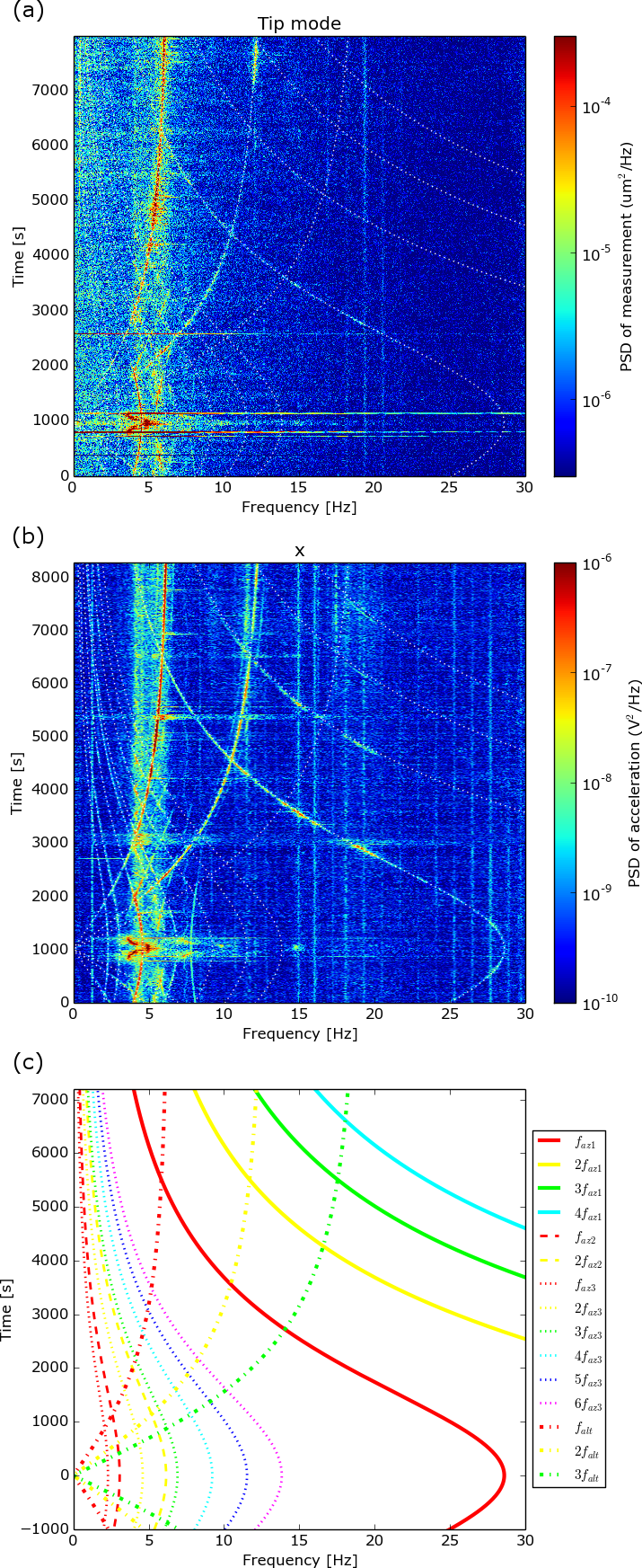}
\end{tabular}
\end{center}
\caption 
{ \label{fig:psdcomp}
Comparison between (a) PyWFS measurements, (b) accelerometer measurements and (c) vibration frequency simulations, for the target GJ 504, with a declination $\delta=9^o19'57"$. The simulations are also overlapped with the measurements with dotted lines, to show the frequency match.}
\end{figure}

During the observation of the brown dwarf companion GJ 504 b previously discovered at the Subaru Telescope\cite{Kuzuhara2013}, accelerometer data were recorded at the same time as pseudo open-loop data from the PyWFS. The PyWFS data were binned to match the accelerometer sampling rate of 60~Hz. Using the frequency calculations from Sec.~\ref{sec:speed}, the vibrations in altitude and azimuth were simulated with their harmonics, and superimposed on the data. Figure~\ref{fig:psdcomp} presents the result of the analysis.

In this figure, a strong correlation between PyWFS and accelerometer data can be seen, in frequency and amplitude. This result shows that all the low-frequency vibrations that affect the PSF in the \sce instrument originate from the telescope motion. The band of amplification between 4 and 6~Hz is also present in the PyWFS data, as well as the strong fast-varying vibration around transit, and some of the constant vibrations around 20~Hz.

The simulated frequencies in altitude $f_\mathrm{alt}=f_{a,\mathrm{inc}}$ and azimuth $f_\mathrm{az1}=f_{A,\mathrm{inc}}$, calculated with the pitches of the incremental encoders, fit perfectly with the measured vibrations. The $f_\mathrm{az1}$ vibration seems to have a low impact and is not amplified around 5~Hz. Harmonics up to the fourth order are also present, although with a lower amplitude.

The $f_\mathrm{alt}$ vibration is much stronger and seems to be the dominant source of disturbance far from transit, for $t>3000$~s. The second harmonics is also very significant, while the third harmonics is only marginally measurable. Although it is not obvious in this example, several measurements at different declinations ---for example the result presented in Fig.~\ref{fig:psdaccel}--- showed that these vibrations are amplified independently from the 4--6~Hz region.

Other components proportional to $f_\mathrm{az1}$ are also noticeable at lower frequency, but unlike the $f_\mathrm{az1}$ vibration and its harmonics, some of these vibrations are amplified in the 4--6~Hz range. One of them, labeled $f_\mathrm{az2}=f_{A,\mathrm{abs}}$, seems to be induced by the pitch of the azimuth absolute encoder. The second harmonic is also  present. Other vibrations not explained by either the incremental nor the absolute encoders are also visible. The main component is defined as
\begin{equation}
	f_\mathrm{az3} = \frac{f_\mathrm{az1}}{12.5} \;\textrm{or}\; f_\mathrm{az3} = 0.75*f_\mathrm{az2}\, ,
\end{equation}
with measured harmonics up to the sixth order. Most of them are never significant enough to deteriorate the PSF quality, but in the example presented in Fig.~\ref{fig:psdcomp}, the second harmonics of this vibration goes through the 4--6~Hz range and gets amplified by almost two orders of magnitude. It is unclear what the origin of these vibrations is, since they cannot be explained by the pitches of the different encoders, but it is possible they are generated by the motors driving the azimuthal rotation of the telescope. An interesting symmetry in frequency of these vibrations around 5~Hz is noticed in both data sets, and will be discussed in the next section.

Smaller vibrations at fixed frequencies can also be noted in both measurements, especially in the accelerometer measurements. They form a regular grid, with a separation of 1.25~Hz. A fixed vibration at the frequency $f=1.25$~Hz is also visible. In the PyWFS data, only a few harmonics around 20~Hz are above the noise level, but match the frequencies measured with the accelerometer. The origin of this regular pattern is unknown, but could originate from the regular pulsation of a mechanism on the telescope. These vibrations have low amplitudes, therefore should not impact the stability of the image.

\subsection{Symmetries Around 5~Hz}
\label{sec:symmetries}

\begin{figure}
\begin{center}
\begin{tabular}{c}
\includegraphics[width=0.5\columnwidth]{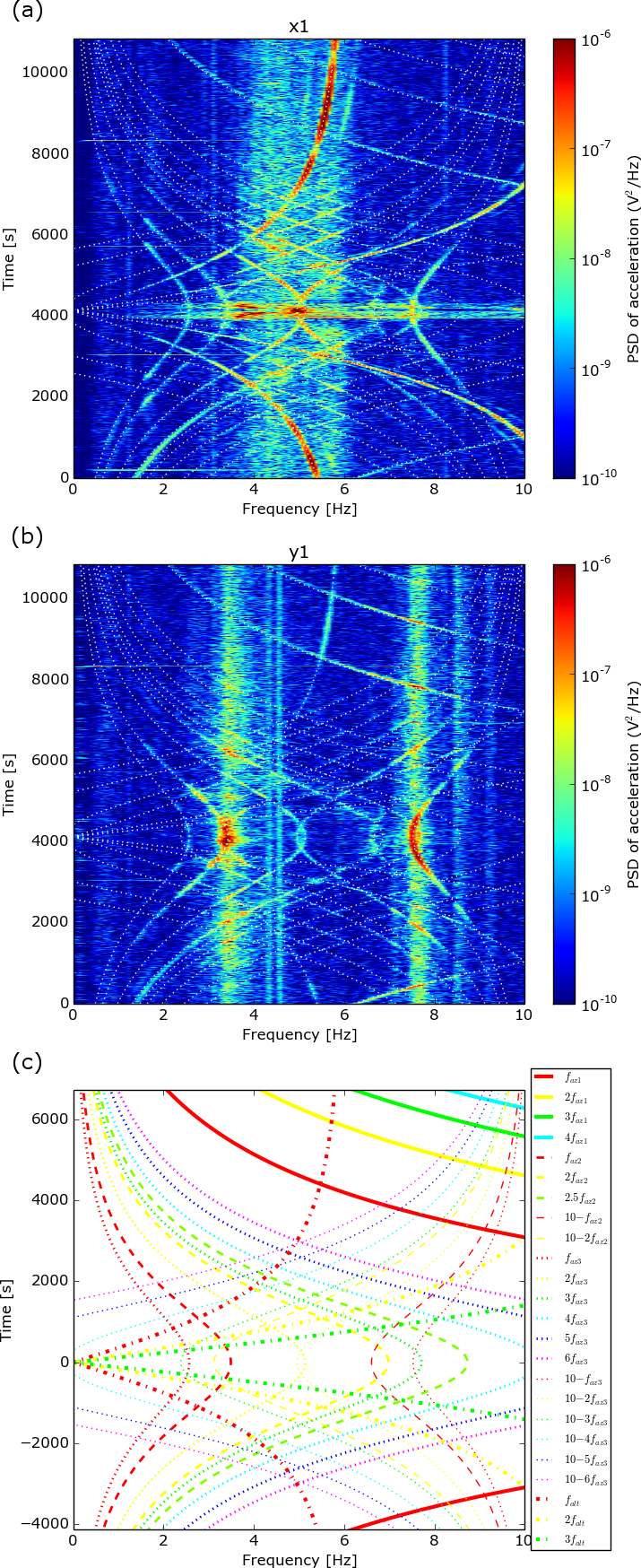}
\end{tabular}
\end{center}
\caption 
{ \label{fig:psdzoom}
Comparison between accelerometer measurements in (a) X and (b) Y, and (c) vibration frequency simulations, for the target \textbeta\xspace Pegasi, with a declination $\delta=28^o10'27"$.}
\end{figure}

Some vibrations at low frequencies, especially in the 4--6~Hz amplification zone, show  a symmetric component around 5~Hz. They can therefore be fitted with a vibration shifted by 10~Hz, twice the symmetric value, with frequencies at $10-Nf_\mathrm{az2}$, with $N\in \left[1,2\right]$, or $10-Nf_\mathrm{az3}$, with $N\in \left[1,6\right]$. Figure~\ref{fig:psdzoom} presents an example of accelerometer measurements on the target \textbeta\xspace Pegasi taken during the \sce engineering night of August 12, 2016. The X axis is reduced to the 0--10~Hz range to see the symmetry more clearly, but the data were acquired at 60~Hz. 

Both measurements show symmetric vibrations around 5~Hz, for all the harmonics of $f_\mathrm{az3}$ and up to the second harmonics of $f_\mathrm{az2}$. Most of these vibrations only appear in the 4--6~Hz amplification region for the X axis, and in the amplification regions around 7.5~Hz and 3.75~Hz for the Y axis. It is interesting to note that the amplification zone measured in the X axis is centered around 5~Hz like the symmetry, indicating a probable common source, namely the azimuthal control loop filter. It is unclear how such a symmetry is created, but since the control loop frequency of the telescope rotation is 10~Hz, it is probably coming from aliasing of the existing vibrations by this control loop.

\begin{figure}
\begin{center}
\begin{tabular}{c}
\includegraphics[width=0.5\columnwidth]{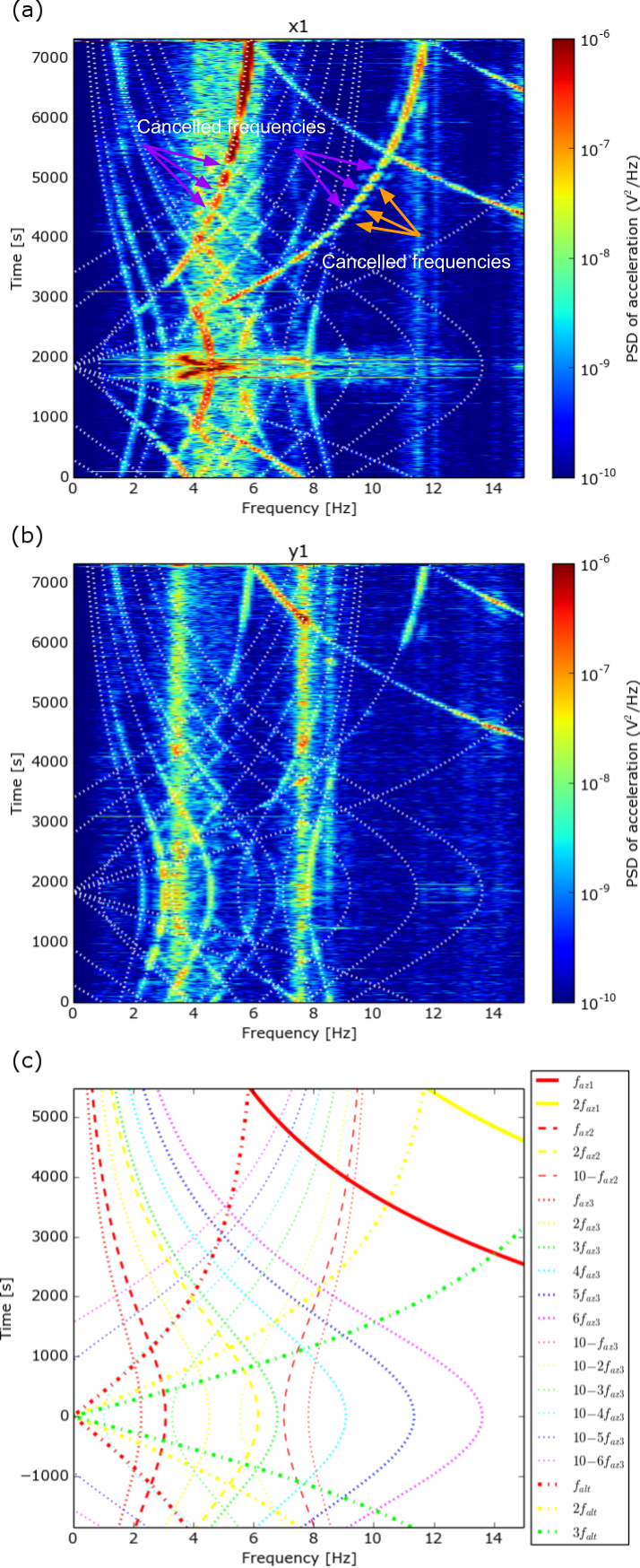}
\end{tabular}
\end{center}
\caption 
{ \label{fig:psdgj504}
Comparison between accelerometer measurements in (a) X and (b) Y, and (c) vibration frequency simulations, for the target GJ504, with a declination $\delta=9^o19'57"$.}
\end{figure} 

Figure~\ref{fig:psdgj504} presents a second example of that symmetry, during the observation of GJ 504 fourteen months after the data were taken for Fig.~\ref{fig:psdcomp}. The accelerometer data were sampled at 120~Hz this time, but the X axis was reduced to the 0--15~Hz range to see the symmetry. The amplitude of the vibrations are similar to the previous observations. An interesting fact to notice here is that the vibration in altitude has a decreased amplitude every time its frequency $f_\mathrm{alt}$ is equal to one of the harmonics of $f_\mathrm{az3}$ (purple arrows on Fig.~\ref{fig:psdgj504} (a)). The second harmonics of $f_\mathrm{alt}$ has the same variations at these frequencies, but also between the nulls of $f_\mathrm{alt}$ (orange arrows on on Fig.~\ref{fig:psdgj504} (a)). We can see the same effect in the LLOWFS measurements presented in Fig.~\ref{fig:psdlowfs}. This seems to be a mechanical effect, where vibrations in azimuth alter and even cancel vibrations in altitude for short periods of time. 

These examples show that, even if it is possible to predict the frequencies of the vibrations using only the coordinates of the target, it is difficult to determine which vibration will be dominant at any given time. 

\subsection{Transit Vibration}
\label{sec:transit}

\begin{figure}
\begin{center}
\begin{tabular}{c}
\includegraphics[width=0.5\columnwidth]{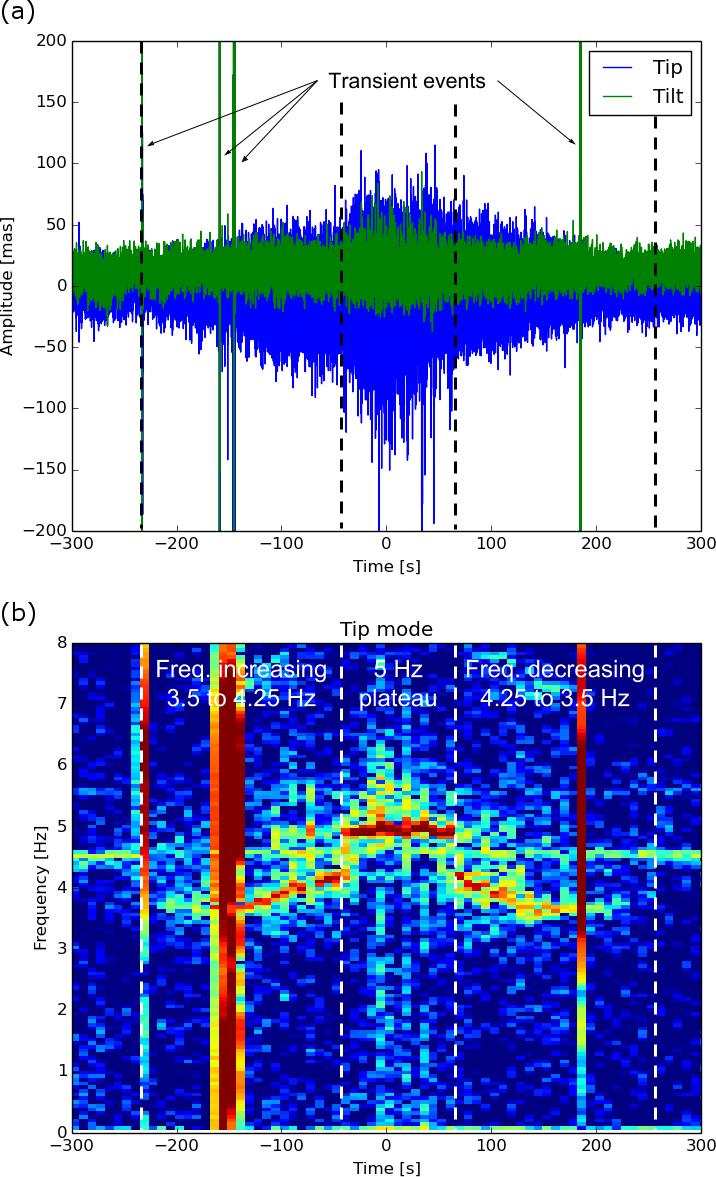}
\end{tabular}
\end{center}
\caption 
{ \label{fig:temptransit}
Evolution of the vibrations in pseudo open-loop with the PyWFS measured around the time of transit, (a) in amplitude (temporal measurement) and (b) in frequency (evolution of the PSD). The loop of AO188 was still closed, correcting part of the vibrations already. The X axis is centered at transit.}
\end{figure}

In all the vibration measurements presented so far, the predominant feature is a strong vibration around the time of transit, with a frequency variation that is not explained by the rotation speed of the telescope. As seen if Figures~\ref{fig:psdlowfs} and \ref{fig:psdpywfs}, the frequency is first constant at about 3.5~Hz, then increases rapidly to about 4.25~Hz. Then it jumps to about 5~Hz, and usually stays at that frequency during transit, before decreasing back to a value between 4.25 and 3.5~Hz. We observed on some occasions the frequency jumping back and forth between 4.25 and 5~Hz instead of staying at 5~Hz. Figure~\ref{fig:temptransit} presents the pseudo open-loop measurement taken with \sce's PyWFS around the time of transit during the observation of GJ 504 presented in Fig.~\ref{fig:psdcomp}.

\begin{figure}
\begin{center}
\begin{tabular}{c}
\includegraphics[width=0.5\columnwidth]{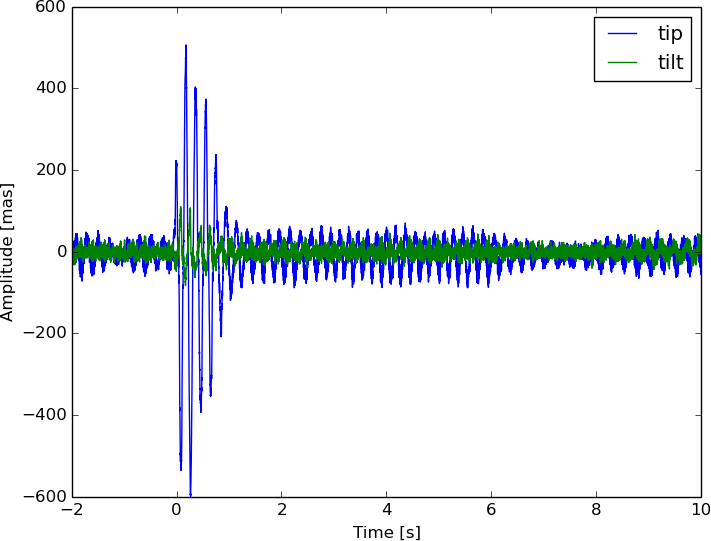}
\end{tabular}
\end{center}
\caption 
{ \label{fig:tempjump}
Example of a transient high-amplitude jump around transit. The origin of the X axis is at the beginning of the jump.}
\end{figure}

As we can see on this figure, the amplitude of the vibration is also increasing when the frequency changes from 3.5 to 4.25~Hz, then plateaus at a frequency of 5~Hz at about 200~mas peak-to-valley ---2.5~times the size of the PSF in H-band---, then decreases again with the frequency. Some transient high-amplitude jumps are also visible during that period. Since they are also measured with the accelerometers, we are confident that it is not an instability in the control loops of AO188 or \sce, but an actual motion of the telescope. Figure~\ref{fig:tempjump} presents an example of such a transient event on a smaller timescale.

This figure shows that the amplitude increases dramatically to more than 1~arcsec peak-to-valley, for only a second, then settles down after five more seconds. The frequency of the vibration does not change during the event, only the amplitude gets affected. These events seem to appear randomly only when the frequency increases or decreases between 4 and 4.5~Hz, without any periodicity. One likely cause is from the telescope cables rotating less smoothly than the telescope itself. Indeed, All the cables reaching the telescope and the dome, are going through a rotating system under the telescope. This system is designed to avoid any tangling during rotation, but it may not rotate as smoothly as the telescope, especially during faster azimuthal rotations. Although this theory does not explain why this effect only appears during the transit vibration.

\begin{figure}
\begin{center}
\begin{tabular}{c}
\includegraphics[width=0.5\columnwidth]{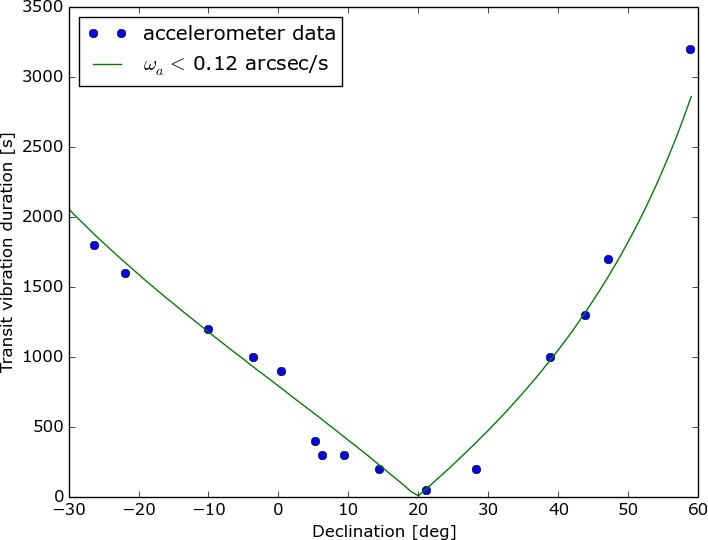}
\end{tabular}
\end{center}
\caption 
{ \label{fig:transittime}
Measurements of the duration of the transit vibration (from the appearance of the vibration to its disappearance), for different targets at various declinations. We also simulated the duration over which the telescope's rotation in altitude is below 0.12~arcsec/s, which fits the measurements.}
\end{figure}

From the various examples presented previously, it is clear that the duration of the transit vibration is dependent of the altitude of the telescope at transit: the closer to zenith the target is, the shorter the vibration will last. Though, its duration seems to be linked to the speed more than the position of the telescope. Figure~\ref{fig:transittime} presents measurements of the duration of the vibration, from the beginning to the end of the events, for targets with different declinations. These measurements were only accurate to about a minute, since it is difficult to estimate precisely the duration of the transit vibration.

This figure shows that the transit vibration can degrade science data for an important part of the observation (i.e. the transit where the most field rotation and hence diversity are acquired). For example, a target with a declination of $\delta = 60^o$ will spend almost an hour in the major vibration. In the extreme case of Polaris ($\delta = 89^o16'$), the telescope is almost static all the time, and the vibration is always present. The measurements presented in Fig.~\ref{fig:transittime} are best fitted by the duration over which the rotation speed of the telescope in altitude $\omega_a$ is below 0.12~arcsec/s, around transit. This indicates that the transit vibration could come from the incremental servo loop (or the current loop presented in Fig.~\ref{fig:loop}), that controls the speed of the telescope, and could lose precision when the speed is too low.

The correction of the transit vibration is one of the main challenges faced by the ExAO system. Contrary to the vibrations linked to the encoders, the high variability in frequency and amplitude of the transit vibration makes it hard to correct it even with more advanced predictive control algorithms like the Linear Quadratic Gaussian (LQG) controller\cite{Lozi2016a}. The transient events are particularly hard to correct with a classical control loop. The amplitude of actuation necessary to correct them is too high for \sce's deformable mirror. But with new algorithms adapting to recurring transient events\cite{Guyon2017,Males2018}, it is possible to compensate for them, especially by using the higher stroke of AO188's deformable mirror. 

\subsection{Extreme Conditions}
\label{sec:extreme}

\subsubsection{Low transit altitude}
\label{sec:lowalt}

\begin{figure}
\begin{center}
\begin{tabular}{c}
\includegraphics[width=0.5\columnwidth]{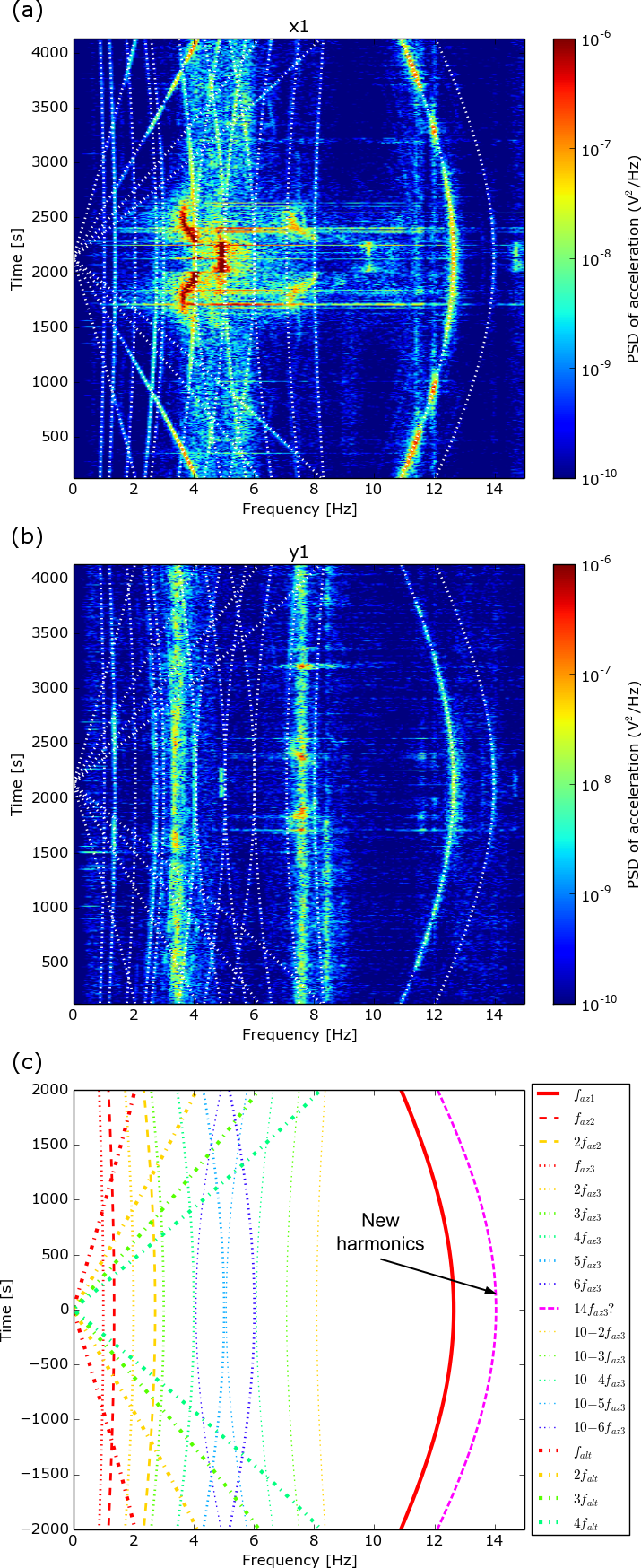}
\end{tabular}
\end{center}
\caption 
{ \label{fig:psdvega}
Comparison between accelerometer measurements in (a) X and (b) Y, and (c) vibration frequency simulations, for the target Vega, with a declination $\delta=38^o48'12"$.}
\end{figure}

The vibration simulations presented in Fig.~\ref{fig:vibaltazvsdel} show that depending on the declination of the target, vibrations can have vastly different impacts on the image quality. For targets transiting far from zenith, vibrations are varying slowly over the course of the observation, but are impacted to a greater extent by the strong transit vibration. Figure~\ref{fig:psdvega} presents accelerometer measurements obtained at 120~Hz on the target Vega, transiting at an altitude of about $70^o$.

In this case, only the transit vibration around 5~Hz is significant at the time of transit. The vibrations from the azimuthal encoder evolve only over a few Hertz around 13~Hz, and none of the harmonics are significantly amplified between 4 and 6~Hz. The altitude vibrations are also small, and only the second harmonics starts to become significant around 4~Hz, at the end of the observation sequence. We note that a new harmonic appeared for this target, around 14~Hz, maybe the 14\textsuperscript{th} order of $f_\mathrm{az3}$. But once again, it's amplitude is very low.

\subsubsection{Transiting around zenith}
\label{sec:highalt}

\begin{figure}
\begin{center}
\begin{tabular}{c}
\includegraphics[width=0.5\columnwidth]{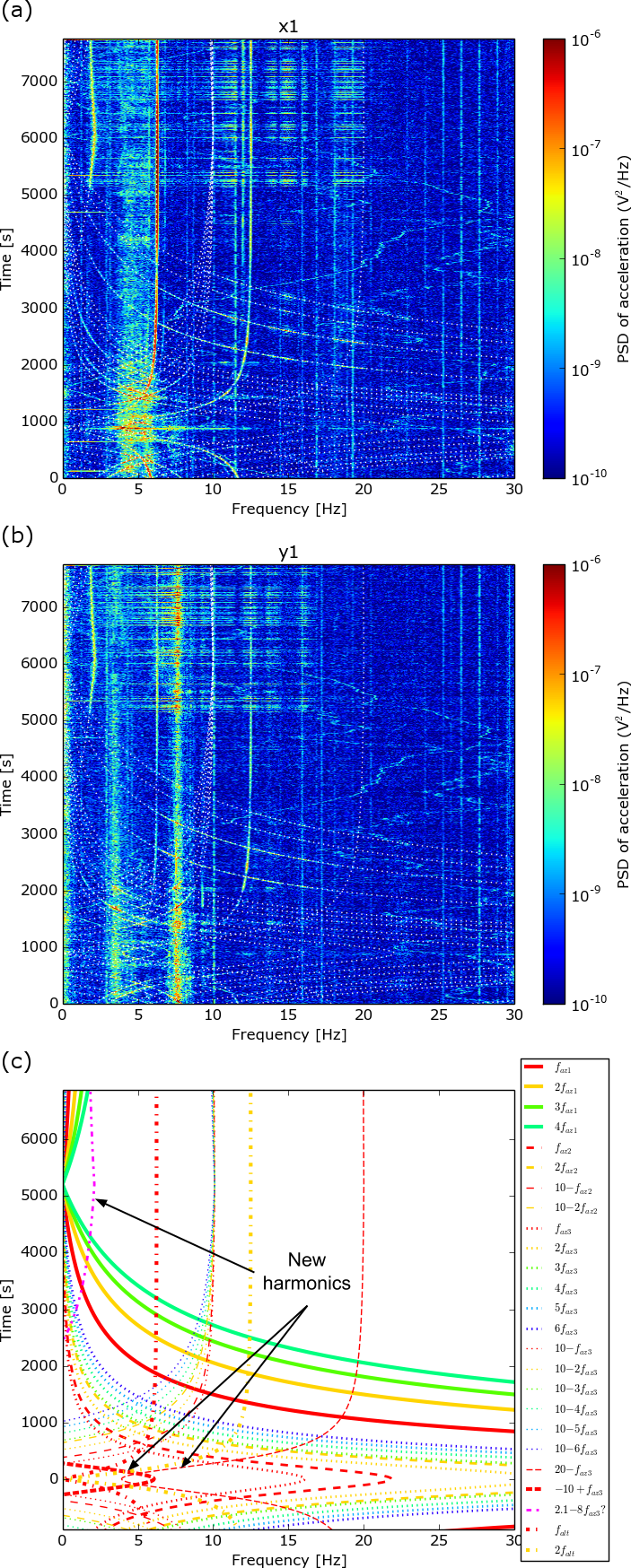}
\end{tabular}
\end{center}
\caption 
{ \label{fig:psdhr8799}
Comparison between accelerometer measurements in (a) X and (b) Y, and (c) vibration frequency simulations, for the target HR 8799, with a declination $\delta=21^o13'29"$.}
\end{figure}

One of the most important targets for high-contrast imaging is the system of four planets around HR 8799\cite{Marois2008}, the only confirmed multi-planet system imaged for now. This system is a benchmark for characterizing contrast and spectral sensitivity of high-contrast instruments. But with a declination at around $21^o$, the system transits at just over one degree from zenith from Maunakea, close to the mechanical limit of the Subaru Telescope. Figure~\ref{fig:psdhr8799} presents accelerometer measurements for HR 8799. 

As expected, the vibrations evolve rapidly around the time of transit, which make them difficult to correct efficiently. The vibrations are changing frequencies so much that new harmonics appear: similarly to the frequencies presented in Sec.~\ref{sec:symmetries}, a vibration at a frequency of $f_\mathrm{az3}-10$ is now visible around 5~Hz, as well as a symmetric vibration at a frequency of $20-f\mathrm{az3}$. 

Another vibration is noticeable around the time the azimuthal rotation slows down and even reverses, at about $t=5000$~s after transit. That vibration might be a symmetric to the eighth order of $f_\mathrm{az3}$. It seems to create transient events similar to the ones observed during transit, although with lower amplitudes. These events might be due to the fact that the azimuthal rotation is under a certain speed, in a similar way that the transit vibration is created when the rotation speed in altitude is lower than 0.12~arcsec/sec.

In this example, the transit vibration only lasts about a minute. But the high variability of the vibration frequencies and amplitudes around transit make this target a difficult one to observe, and it requires a system that can adapt to fast variations.

\section{CONCLUSION}
\label{sec:conclusion}

After atmospheric turbulence, telescope vibrations are the strongest disturbance that high-contrast imaging instruments like \sce face. Despite the fact that they tend to be at low frequencies due to the mass of the telescope, their amplitudes make them hard to correct, even when the control loop runs at a few kHz.

In this paper, we demonstrated that various wavefront sensors inside \sce see the same type of vibrations, evolving in frequency and amplitude with time. Using accelerometers installed on the top ring supporting the secondary mirror of the telescope, we were able to measure the same vibrations, and a comparison between wavefront sensing measurements and accelerometer measurements show a strong correlation in frequency and amplitude.

The slow-varying vibrations are proportional to the speed of the telescope in altitude and azimuth. Most of them can be simulated using the frequencies at which the various encoders are read, as well as their harmonics. Some vibrations are amplified at fixed frequencies, especially in a range between 4 and 6~Hz. Symmetric vibrations are also created in that range, probably from the telescope drive control. Being able to predict the frequencies of the vibrations is an important step in the development of an optimal controller\cite{Meimon2010}.

The main difficulty in the optimal control of the telescope vibrations is the highly unpredictable vibrations around the time of transit. That time is crucial for high-contrast imaging, because it provides most of the field rotation necessary for post-processing algorithms like angular differential imaging. That vibration is usually strong, its frequency is harder to simulate, and high-amplitude transient events are present. These event are especially hard to correct properly since they are short and random.

This study provides useful informations concerning the origins and impacts of the telescope vibrations. Actions can now be taken to correct some of these effects directly on the telescope, for example by optimizing the control loop of the pointing. But since modifying things at the telescope level is complicated, new wavefront control techniques are being studied. An adaptive Linear Quadratic Gaussian (LQG) controller was implemented and showed promising results\cite{Lozi2016a}, while a more robust predictive control is also currently tested on \sce\cite{Guyon2017,Males2018}. A increased communication between AO188 and \sce will soon help the correction of strong vibrations, by offloading part of the correction to AO188's higher stroke DM. Finally, future techniques like sensor fusion, which merges measurements from multiple wavefront sensors and accelerometers to compute an optimal correction, will also be implemented in the future, and should improve the correction of strong vibrations. The knowledge of vibrations provided by this work will greatly improve the performance of predictive control algorithms.

\acknowledgments 
 
The development of SCExAO was supported by the Japan Society for the Promotion of Science (Grant-in-Aid for Research \#23340051, \#26220704 \& \#23103002), the Astrobiology Center of the National Institutes of Natural Sciences, Japan, the Mt Cuba Foundation and the directors contingency fund at Subaru Telescope. G. Singh would also like to acknowledge her appointment to the NASA Postdoctoral Program at the Jet Propulsion Laboratory, California Institute of Technology, administered by Universities Space Research Association under contract with NASA. The authors wish to recognize and acknowledge the very significant cultural role and reverence that the summit of Maunakea has always had within the indigenous Hawaiian community.  We are most fortunate to have the opportunity to conduct observations from this mountain.


\bibliographystyle{spiejour}   

\end{document}